\documentclass[aps,amssymb,floatfix,prd,amsmath,preprintnumbers]{revtex4}
\usepackage{epstopdf}
\usepackage{capt-of}
\usepackage{graphicx}  % Include figure files
\usepackage{dcolumn}   % Align table columns on decimal point
\usepackage{bm}% bold math
\begin{document}
\input epsf.tex
%%%%%%%%%%%%
%%%%%%%%%%%

\title{\bf Bianchi-V String Cosmological model with Dark Energy Anisotropy}

\author{B. Mishra\footnote{Department of Mathematics,
Birla Institute of Technology and Science-Pilani, Hyderabad Campus,
Hyderabad-500078, India, email: bivudutta@yahoo.com},  S. K. Tripathy \footnote{Department of Physics, Indira Gandhi Institute of Technology, Sarang, Dhenkanal, Odisha-759146, India, email: tripathy\_sunil@rediffmail.com} and Pratik P. Ray \footnote{Department of Mathematics,
Birla Institute of Technology and Science-Pilani, Hyderabad Campus,
Hyderabad-500078, India, email: pratik.chika9876@gmail.com}
}

\affiliation{ }

\begin{abstract}
\begin{center}
\textbf{Abstract}
\end{center}
The role of anisotropic components on the dark energy and the dynamics of the universe is investigated. An anisotropic dark energy fluid with different pressures along different spatial directions is assumed to incorporate the effect of anisotropy. One dimensional cosmic strings aligned along x-direction supplement some kind of anisotropy. Anisotropy in the dark energy pressure is found to evolve with cosmic expansion at least at late times. At an early phase, the anisotropic effect due to the cosmic strings substantially affect the dynamics of the accelerating universe.
\end{abstract}

\maketitle
%\textbf{PACS number}:\\

\section{Introduction}
It has now become an accepted fact that the expansion of the universe is accelerating due to an exotic energy source having large negative pressure called dark energy (DE). But the mystery is that, a little is known about dark energy: it violates strong energy condition and can cluster at large scale.  Moreover DE does not interact with baryonic matter and therefore making it difficult to detect. It dominates the present universe and was less effective in early time. Recent Planck results estimate a lion share of $68\%$ for DE in the cosmic mass energy budget \cite{Ade14}. The natural choice for DE is the cosmological constant but it suffers from many puzzles like the fine tuning and coincidence problem. Therefore many dynamically varying DE candidates have been proposed in literature (for details see Refs.\cite{Copeland, Wang16} for reviews). Observations have confirmed that the cosmic speed up is a late time phenomena and has occurred at a redshift of the order $z_t \sim 1$. This indicates that, the universe has undergone a transition from a decelerated phase of expansion to an accelerating state in recent past. This cosmic transit phenomenon speculates an evolving deceleration parameter with a signature flipping. The rate at which the transition occurs usually determines the transit redshift $z_t$.

The standard cosmological model is based upon the assumption of large scale isotropy and  homogeneity of space.  However, one can expect small scale anisotropies in the universe.  There have been a lot of investigations on the anisotorpic features of the universe in recent times \cite{Watan09, Buiny06,tripa2, Ade14, Quartin2015, Ade16, Picon04, Jaffe05, Koivisto08a, Cooray10, Cook10,Colin11, Salehi16, campa1, campa3, gruppo, Koivisto06, Koivisto08, Bat09, akarsu1, akarsu2, Sadeh16, Bunn96}. Using the data from WMAP, many authors have claimed a departure from global isotropy so that there occurs a non trivial topology of the large scale cosmic geometry with asymmetric expansion \cite{ Watan09, Buiny06,tripa2}. The preliminary Planck data show a slight redshit of the power spectrum from exact scale invariance \cite{Ade14} eventhough some other analysis of Planck data go against the global anisotropy \cite{Quartin2015, Ade16}. Dark energy is also believed to be associated with the breakdown of global isotropy and can display anisotropic features \cite{Picon04, Jaffe05, Koivisto08a, Cooray10}. Employing the Union compilation of Type Ia Supernovae, Cook and Lynden-Bell found that a weak anisotropy in dark energy, mostly observable for higher redshift group with $z>0.56$, directed  roughly towards the cosmic microwave background dipole \cite{Cook10}. Also, Colin et al. showed that the isotropic $\Lambda$CDM model is not consistent with the Union2
data with $z<0.05$ at $2 - 3\sigma$ \cite{Colin11}.  The issue of the departure from global isotropy may be resolved in near future. However, Bianchi type models have been constructed in recent times to handle the issue of anisotropy \cite{campa1, campa3, gruppo, Koivisto06, Koivisto08, Bat09, akarsu1, akarsu2}. It is certain that Bianchi type models are the generalisation of the FRW models where the spatial isotropy is relaxed. 

In the present work, we consider a general form of non interacting dark energy with anisotropic pressures along  different spatial directions and have investigated the effect of anisotropic components on the dark energy density parameter and the equation of state. In addition to the anisotropic DE fluid, cosmic strings aligned along x-direction are also considered to incorporate some anisotropic effect. In some of our earlier studies, we have investigated the dynamical behaviour of pressure anisotropies either in the framework of GR or alternative gravity theory \cite{Mishra15, mishrampla, skt15}. Cosmological studies involving one dimensional cosmic strings are interesting in the sense that, they may provide some explanation to the Physics of the early universe. Moreover, cosmic strings are believed to arise as topological defects during the spontaneous symmetry breaking. They may help in the formation of large structures like galaxies. There are some interesting investigations on different aspects of string cosmological models in recent times \cite{SKT08, SKT09a, SKT09b,SKT09c, Saha08}.

The paper is organized as follows. In Sect. 2, we discuss the basic formalism of an anisotropic dark energy. The effect of anisotropy on dark energy and the DE equation of state (EoS) parameter is discussed in Sect. 3 and we summarize our results in Sect. 4.
 
\section{Basic Formalism}
We consider a spatially homogeneous and anisotropic Bianchi V (BV) space time in the form

\begin{equation} \label{eq:1}
ds^{2}= dt^{2}-a_1^2 dx^2-e^{2\alpha x}\left(a_2^2 dy^2+a_3^2 dz^2\right)
\end{equation}
where, the directional scale factors $a_i (i=1,2,3)$ are functions of cosmic time only, $a_i=a_i(t)$. In general, $a_i$s are considered to be different and thereby provide a description for anisotropic expansions along the three orthogonal spatial directions. The model reduces to FRW model when the directional scale factors become equal. Here, $\alpha$ is a non zero arbitrary positive constant. We choose the unit system such that $8\pi G=c=1$, $G$ is the Newtonian gravitational constant and $c$ is the speed of light. The energy momentum tensor for a given environment of two non interacting fluids is given by

\begin{equation}
T_{\mu\nu}=T_{\mu\nu}^{s}+T_{\mu\nu}^{D} \label{eq:2}
\end{equation}
where $T_{\mu\nu}^{s}$ and $T_{\mu\nu}^{D}$ respectively denote the contribution to energy momentum tensor from one dimensional cosmic strings and DE. For a cosmic fluid containing one dimensional strings \cite{letelier, stachel}, the energy momentum tensor is 

\begin{equation}\label{eq:3} 
T^{s}_{\mu\nu} = (\rho + p) u_{\mu} u_{\nu}- pg_{\mu\nu}+ \lambda x_{\mu}x_{\nu}
\end{equation}
where $u^{\mu}u_{\mu}=-x^{\mu}x_{\mu}=1$ and $u^{\mu}x_{\mu}=0$. In a co moving coordinate system, $u^{\mu}$ is the four velocity vector and $p$ is the isotropic pressure of the fluid. $x^{\mu}$ represents the direction of cosmic strings (here x-direction). $\rho$ is the proper energy density and is composed of energy density due to massive particles and the string tension density $\lambda$. In the absence of any string phase, the total contribution to the baryonic energy density comes from particles only. In contrast to the isotropic pressure of usual cosmic fluid, we wish to incorporate some degree of anisotropy in the DE pressure and consider the energy momentum tensor for DE as

\begin{eqnarray} \label{eq:4}
T^{D}_{\mu\nu} & = & diag[-\rho_{D}, p_{D_x},p_{D_y},p_{D_z}] \nonumber\\ 
			 % & = & diag[-1, \omega_{de_x},\omega_{de_y},\omega_{de_z}]\rho_{de}\\ \nonumber
			  & = & diag[-1, \omega_{D}+\delta,\omega_{D}+\gamma,\omega_{D}+\eta]\rho_{D} ,
\end{eqnarray}
where $\omega_{D}$ is the DE equation of state parameter and $\rho_{D}$ is the DE density. The skewness parameters $\delta$, $\gamma$ and $\eta$ are the deviations from $\omega_{D}$ along $x$, $y$ and $z$ axes respectively. The DE pressure becomes isotropic when $\delta$, $\gamma$ and $\eta$ vanish identically.

The field equations, $G_{\mu\nu}=T_{\mu\nu}$, for a two fluid system consisting of cosmic strings and DE in a BV metric are obtained as  

\begin{equation}\label{eq:5}
\frac{\ddot{a_2}}{a_2}+\frac{\ddot{a_3}}{a_3}+\frac{\dot{a_2}\dot{a_3}}{a_2a_3}-\frac{\alpha^{2}}{a_1^{2}}=-p-\lambda-(\omega_{D}+\delta)\rho_{D}  
\end{equation} 

\begin{equation}\label{eq:6}
\frac{\ddot{a_1}}{a_1}+\frac{\ddot{a_3}}{a_3}+\frac{\dot{a_1}\dot{a_3}}{a_1a_3}-\frac{\alpha^{2}}{a_1^{2}}=-p-(\omega_{D}+\gamma)\rho_{D} 
\end{equation}

\begin{equation}\label{eq:7}
\frac{\ddot{a_1}}{a_1}+\frac{\ddot{a_2}}{a_2}+\frac{\dot{a_1}\dot{a_2}}{a_1a_2}-\frac{\alpha^{2}}{a_1^{2}}=-p-(\omega_{D}+\eta)\rho_{D}
\end{equation}

\begin{equation}\label{eq:8}
\frac{\dot{a_1}\dot{a_2}}{a_1a_2}+\frac{\dot{a_2}\dot{a_3}}{a_2a_3}+\frac{\dot{a_3}\dot{a_1}}{a_3a_1}-\frac{3\alpha^{2}}{a_1^{2}}=\rho+\rho_{D} 
\end{equation}

\begin{equation} \label{eq:9}
2\frac{\dot {a_1}}{a_1}-\frac{\dot{a_2}}{a_2}-\frac{\dot{a_3}}{a_3}=0,
\end{equation}
where an over dot represents the differentiation with respect to time $t$. One can note that, anisotropy is incorporated in two different manner in the field equations: one due to the inherent anisotropic DE fluid and the other through the presence of aligned cosmic strings along x-axis. These anisotropic components leads to the anisotropy in the cosmic geometry described through the BV space time.

The cosmic spatial volume and the scalar expansion $\theta$ can be obtained respectively as $V=R^3=a_1a_2a_3$ and $\theta=3H$, where $H=\frac{\dot{R}}{R}=\frac{1}{3}\sum_{i=1}^{3} H_i$ is the mean Hubble parameter.  $H_i=\frac{\dot{a_i}}{a_i}$ are the directional Hubble parameters along different spatial directions. The shear scalar is given by $\sigma^2=\frac{1}{2}(\sum_{i=1}^{3} H_i^2-\frac{\theta^2}{3})$ and the deceleration parameter can be calculated from the relation $ q=\frac{d}{dt} \left(H^{-1}\right)-1$. 

It is straightforward to obtain $a_1^2=a_2a_3$ from \eqref{eq:9}. In order to get an anisotropic relation between the directional scale factors, we assume 
\begin{equation}
a_2=a_3^k,
\end{equation}
where $k$ is an arbitrary positive constant else than 1. Consequently, the scale factors along different directions can be $a_1=R,~ a_2=R^{\frac{2k}{k+1}}$ and $a_3=R^{\frac{2}{k+1}}$. The directional Hubble parameters can be obtained as, $H_1=H$, $H_2=\big(\frac{2k}{k+1}\big)H$ and $H_3=\big(\frac{2}{k+1}\big)H$.  The anisotropic parameter $\mathcal{A}$ can be obtained as 
\begin{equation}
{\mathcal A}=\frac{1}{3}\sum \left(1-\frac{H_i}{H}\right)^2=\frac{2}{3}\left(\frac{k-1}{k+1}\right)^2.
\end{equation}
The anisotropic nature of model can also be  quantified through the estimation of $\frac{\sigma}{H}$ at the present epoch. In many earlier works, there have been discussions on this quantity which may come out to be constant with time \cite{SKT10, Muh18,Saha12,Collins77, Collins80, Roy85, Roy95, Bali08, Sharif10}. For the BV model, this quantity can be obtained as 

\begin{equation}
\frac{\sigma}{H}= \left[\frac{5k^2+2k+5}{2(k+1)^2}-1.5\right]^{\frac{1}{2}}.\label{eq:15a}
\end{equation}
Observational limits to this quantity have been obtained by many workers in recent times. From an analysis of COBE data, Bunn et al. have obtained a limit of $\left(\frac{\sigma}{H}\right)_0 < 3 \times 10^{-9}$ \cite{Bunn96}. Saadeh et al. put a tighter constraint of  $\left(\frac{\sigma}{H}\right)_0 < 4.7 \times 10^{-11}$ which strongly disfavours the anisotropic expansion of the universe \cite{Sadeh16}. We may constrain the parameter $k$ according to these bounds. In the present work, we have formulated a model, which can be suited better for any amount of cosmic anisotropy. However, we use the previous constraint on $k$ i.e $k=1.0001633$ so that  $\left(\frac{\sigma}{H}\right)_0$ becomes $8.164 \times 10^{-5}$ \cite{skt15} to obtain some considerable effects on the dynamical features of the universe.

The energy conservation equation for the anisotropic fluid, $T^{\mu\nu}_{;\nu}=0$ yields 

\begin{equation} \label{eq:10}
\dot{\rho}+3(p+\rho)H+\lambda H_1+\dot{\rho_{D}}+3\rho_{D}(\omega_{D}+1)H+\rho_{D}(\delta H_1+\gamma H_2+\eta H_3)=0.
\end{equation}

We consider the cosmic string and DE to be non interacting and obtain two separate equations from \eqref{eq:10},

\begin{equation} \label{eq:11}
\dot{\rho}+3H\left(\rho+p+\frac{\lambda}{3}\right)=0
\end{equation}
and
\begin{equation} \label{eq:12}
\dot{\rho_{D}}+3H\rho_{D}(\omega_{D}+1)+\rho_{D}(\delta H_1+\gamma H_2+\eta H_3)=0.
\end{equation}
The equations of state for strings and isotropic fluid can be considered respectively as 
\begin{eqnarray}
\lambda &=& 3\xi \rho,\\
p &=& \omega \rho,
\end{eqnarray}
where $\xi$ and $\omega$ are assumed to be non evolving state parameters. 

From \eqref{eq:11}, we get
\begin{equation} \label{eq:13}
\rho=\rho_0R^{-3(1+\omega+\xi)}
\end{equation}
where $\rho_0$ is rest energy density due to the matter field at the present epoch. Subsequently, pressure and string tension density can be obtained as

\begin{eqnarray} 
p &=& \omega\rho_0R^{-3(1+\omega+\xi)},\label{eq:14}\\
\lambda &=& 3\xi\rho_0R^{-3(1+\omega+\xi)}. \label{eq:15}
\end{eqnarray}

The DE density is obtained from \eqref{eq:8} and \eqref{eq:13} as

\begin{equation}\label{eq:16}
\rho_{D}= 3\left(\Omega_{\sigma}-\Omega_k\right)\left(\frac{\dot{R}}{R}\right)^2-\rho,
\end{equation}
where $\Omega_{\sigma}= 1-\frac{\mathcal{A}}{2}$ and $\Omega_k=\frac{\alpha^{2}}{\dot{R}^{2}}$.
The density parameters can be expressed as
\begin{eqnarray}
\Omega_{D} &=& \Omega_{\sigma}-\Omega_k-\Omega_m, \label{eq:17}\\
\Omega_m &=& \frac{\rho}{3H^2}.\label{eq:18}
\end{eqnarray}
The total density parameter becomes
\begin{equation}
\Omega=\Omega_m+\Omega_{D}= \Omega_{\sigma}-\Omega_k. \label{eq:19}
\end{equation}
Obviously, for a flat isotropic universe with $\alpha=0$ and $k=1$, the total density parameter reduces to be 1. However, in an anisotropic background the dynamics of dark energy density parameter is substantially affected according to the behaviour of $\Omega_{\sigma}$.

Eq. \eqref{eq:12} can be split into two parts corresponding to the deviation free part and the one involving  the deviation from EoS parameter:

\begin{eqnarray} 
\dot{\rho_{D}}+3\rho_{D}(\omega_{D}+1)H & =& 0,\label{eq:10a}\\
\delta H_1+\gamma H_2+\eta H_3 &=&0. \label{eq:10b}
\end{eqnarray}

These equations can be expressed as 
\begin{eqnarray}
\dot{\rho_{D}}+3\rho_{D}(\omega_{D}+1)\frac{\dot{R}}{R} &=& 0,\label{eq:20}\\ 
\left[\delta +\gamma \left(\frac{2k}{k+1}\right)+\eta \left(\frac{2}{k+1}\right)\right]\rho_{D} \frac{\dot{R}}{R} &=& 0. \label{eq:21}
\end{eqnarray}

%Integration of \eqref{eq:20} yields the DE density as
%\begin{equation} \label{eq:22}
%\rho_{D}=\rho_{D_0} R^{-3(\omega_{D}+1)}.
%\end{equation}
%
%It is obvious that, the behaviour of DE density $(\rho_{D})$ is diagnosed by the deviation free part of DE EoS. Here, $\rho_{D_0}$ is the DE density at present epoch. 

We obtain the skewness parameters from the field equations \eqref{eq:5},\eqref{eq:6}and \eqref{eq:7} as
\begin{eqnarray}
\delta &=& -\frac{2}{3\rho_{D}}\left[\zeta^2(k) F(R)+\lambda\right] , \label{eq:23}\\
\gamma &=& \frac{1}{3\rho_{D}}\biggl[\frac{(k+5)}{(k+1)} \zeta(k) F(R)+\lambda\biggr], \label{eq:24}\\
\eta &=& -\frac{1}{3\rho_{D}}\biggl[\frac{(5k+1)}{(k+1)} \zeta(k) F(R)-\lambda\biggr], \label{eq:25}
\end{eqnarray}
where  $\zeta (k)= \dfrac{k-1}{k+1}$, represents the amount of deviation from isotropic behaviour of the model and $F(R)=\biggl( \dfrac{\ddot{R}}{R}+ \dfrac{2 \dot{R}^{2}}{R^{2}} \biggr)$. Besides the control of the parameter $k$, the pressure anisotropies are decided by the behaviour of the functional $F(R)$ and the string tension density. In some of the models in GR, the functional $F(R)$ may vanish leading to an isotropic DE fluid in the absence of any aligned cosmic strings. For isotropic model  with $k=1$, $\zeta(k)$ vanishes and the pressure anisotropies are developed only due to the presence of one dimensional cosmic strings.

The DE EoS parameter $\omega_{D}$ is now obtained as,

\begin{align} \label{eq:26}
- \omega_{D} \rho_{D} & = \left[\dfrac{2\ddot{R}}{ R} +  \left(\dfrac{\dot{R}}{R}\right)^2 \right] \Omega_{\sigma}-\dfrac{\alpha ^{2}}{R^2}+ \rho \left(\omega+ \xi \right) , 
\end{align}

It is interesting to note that, in a cosmic fluid that blends DE with one dimensional cosmic strings in a non interacting manner, the dynamics of the DE EoS parameter will also be governed by the presence of the cosmic string fluid. In the absence of  any matter field, the skewness in pressure anisotropies and the DE EoS reduce to 
\begin{eqnarray} 
\delta &=& -\frac{2}{3\rho_{D}}\biggl[\zeta^2(k) F(R)\biggr] , \label{eq:27}\\
\gamma &=& \frac{1}{3\rho_{D}}\biggl[\frac{(k+5)}{(k+1)} \zeta(k) F(R)\biggr], \label{eq:28}\\
\eta &=& -\frac{1}{3\rho_{D}}\biggl[\frac{(5k+1)}{(k+1)} \zeta(k) F(R)\biggr], \label{eq:29}\\
- \omega_{D} \rho_{D} & =& \left[\dfrac{2\ddot{R}}{ R} +  \left(\dfrac{\dot{R}}{R}\right)^2 \right] \Omega_{\sigma}-\dfrac{\alpha ^{2}}{R^2}. \label{eq:30}
\end{eqnarray}
The above expressions \eqref{eq:27}-\eqref{eq:30} are the same as obtained in our earlier work \cite{mishrampla}. One should note that, the evolutionary behaviour of the pressure anisotropies and the DE EoS depend on the assumed dynamics of the present day universe. In particular, if we have a presumed dynamics of the universe in the form of a scale factor pertaining to the late time cosmic speed up, the background cosmology can be well studied. Consequently, the evolutionary behaviour of these properties can be well assessed. In view of this, in the present work, we consider a hybrid scale factor ( as in Refs. \cite{mishrampla, BM18}) that can simulate a transit universe from a decelerated phase to an accelerated phase.

\section{Model with a Hybrid scale factor}

It has been an usual practice to consider the scale factor to behave either as de Sitter type expansion or  as the power law expansion. Both the power law and exponential law of the scale factor lead to a constant deceleration parameter. However, the belief that the universe has undergone a transition from a decelerated phase of expansion to an accelerated one requires the deceleration parameter to evolve from some positive values at remote past to negative values at late phase of cosmic evolution. Such a behaviour can be generated by a hybrid scale factor \cite{mishrampla, akarsu3, BM18}. The hybrid scale factor behaves as power law in the early cosmic phase and as de Sitter type at late phase of cosmic acceleration. It is expressed through two adjustable parameters $a$ and $b$ as $R=e^{at}t^{b}$. For this model, the Hubble parameter is expressed as $H = a+\dfrac{b}{t}$. Moreover, Tripathy \citep{tripa2} has considered a general form $H = a+ \dfrac{b}{t^{n}}$ of the Hubble parameter to address the role of skewness in anisotropic models where $n$ is a positive constant . The present case can be considered as a special case of that in Ref. \cite{tripa2}. The deceleration parameter for this model is obtained as $q =   -1+ \dfrac{b}{(at+b)^{2}}$. The parameter $b$ can be constrained in the range $0<b<1$ from simple arguments but $a$ can be constrained from a detailed analysis of $H(z)$ data \cite{mishrampla}. In the present work, we consider the same range for $b$ and leave $a$ as an open parameter.

The deceleration parameter generated by a hybrid scale factor is shown in Fig.1. Deceleration parameter for power law cosmology and de Sitter type expansion have also been shown in the figure for comparison. The deceleration parameter for hybrid scale factor evolves from positive values at past epochs to negative values at late time. The transition from deceleration to acceleration occurs at some transition redshift $z_t$ which is decided by the parameter $a$. In the present work, we have considered two representative values of $a$ i.e. $0.1$ and $0.075$ corresponding to $z_t=0.8$ and $0.4$ respectively. It is worth to mention here that, the value of transition redshift has been constrained in a recent work as $z_t=0.806$ \cite{Jesus17}. Also, Farooq et al. obtained similar results from an analysis of Hubble parameter data \cite{Farooq17}. It is clear from the figure that at an early time, the behaviour is dominated by the power law factor and the exponential factor dominates at late times . The rate of transition depends on the parameter $a$; transition is faster for higher value of $a$.

\begin{figure}[h!]
%\minipage{0.40\textwidth}
\includegraphics[scale=0.45]{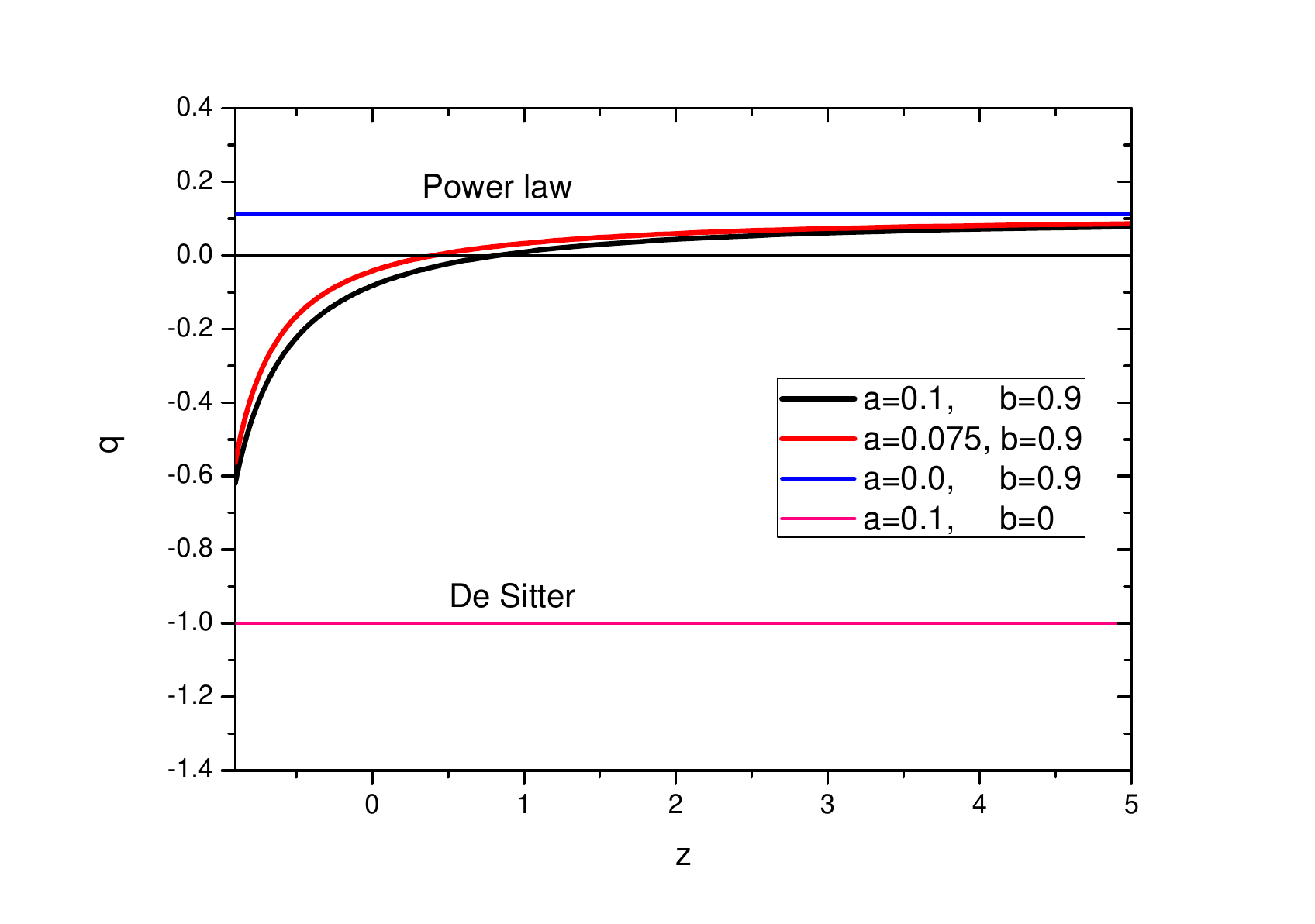}
\caption{Deceleration parameter $q$  as a function of redshift.}
%\endminipage
\end{figure}

%\begin{figure}[h!]
%%\minipage{0.40\textwidth}
%\includegraphics[scale=0.5]{fig4}
%\caption{Variation of Hubble parameter $q$  as a function of redshift.}
%%\endminipage
%\end{figure}
%\begin{figure}[h!]
%%\minipage{0.40\textwidth}
%\includegraphics[scale=0.45]{fig3a}
%\caption{Evolution of skewness parameters in presence and absence of cosmic strings. $\xi$ is taken to be 1/6.}
%%\endminipage
%\end{figure}

\subsection{Dark energy density and skewness parameters}

The DE density parameter has contributions from the anisotropic aspects of the model and the matter field (including the string tension density). For an isotropic model with $k=1$, the anisotropic part $\Omega_{\sigma}$ becomes 1 (refer to eq. \eqref{eq:17}). It is obvious that at remote past, the matter field has a dominant contribution whereas at late phase of cosmic evolution the dark energy dominates. With an increase in the geometrical anisotropy through an increase in the parameter $k$, $\Omega_{\sigma}$ decreases from its isotopic value of 1 but the contributions from matter field and curvature part remain the same. This indicates that, with an increase in the value of $k$, there occurs a decrease in the value $\Omega_D$ at a given redshift. This feature of the DE density parameter is clearly visible in Fig.2, where we have shown $\Omega_D$ as a function of redshift for some representative values of $k$. 

In order to investigate the effect of string phase on DE density parameter, we have considered three different values of $\xi$ namely $\xi=0, \frac{1}{6}$ and $\frac{1}{3}$ corresponding to no string phase, Nambu strings and geometric string phase respectively. Presence of string phase incorporates some kind of anisotropy. It is evident from Fig.3 that the presence of string phase affects the DE density parameter in the early phase of cosmic evolution. With an increase in the value of $\xi$, $\Omega_D$ decreases to some lower values in the past epoch. However, at late times, aligned cosmic strings seem to have little effect on $\Omega_D$. 
%\begin{figure}[h!]
%%\minipage{0.40\textwidth}
%\includegraphics[scale=0.5]{fig3a}
%\caption{Evolution of skewness parameters in presence and absence of cosmic strings. $\xi$ is taken to be 1/6.}
%%\endminipage
%\end{figure}

\begin{figure}[t]
%\minipage{0.40\textwidth}
\includegraphics[scale=0.45]{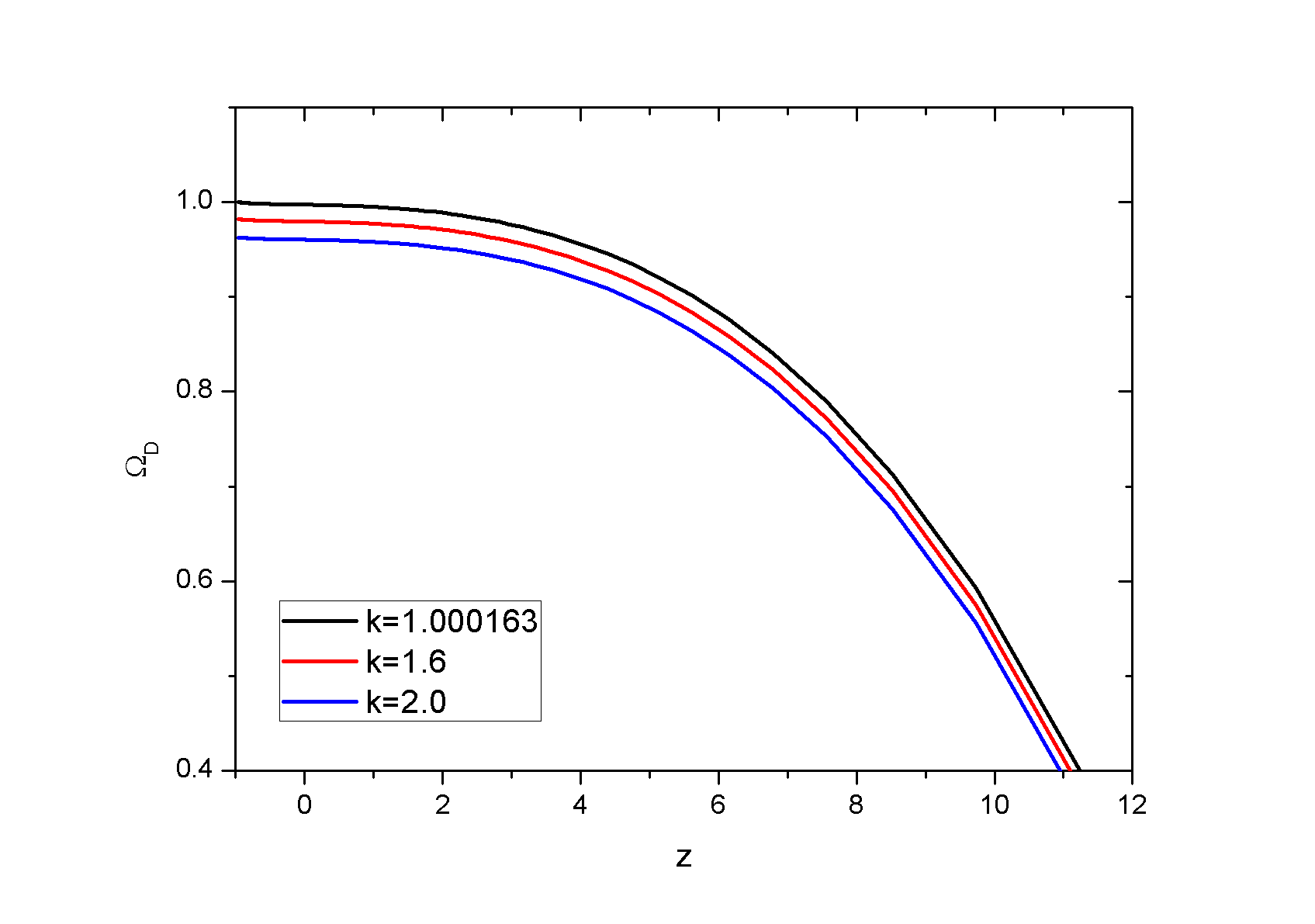}
\caption{DE density parameter as a function of redshift for three representative values of $k$. $\xi$ is taken to be 1/6.}
%\endminipage
\end{figure}

\begin{figure}[h]
%\minipage{0.40\textwidth}
\includegraphics[scale=0.45]{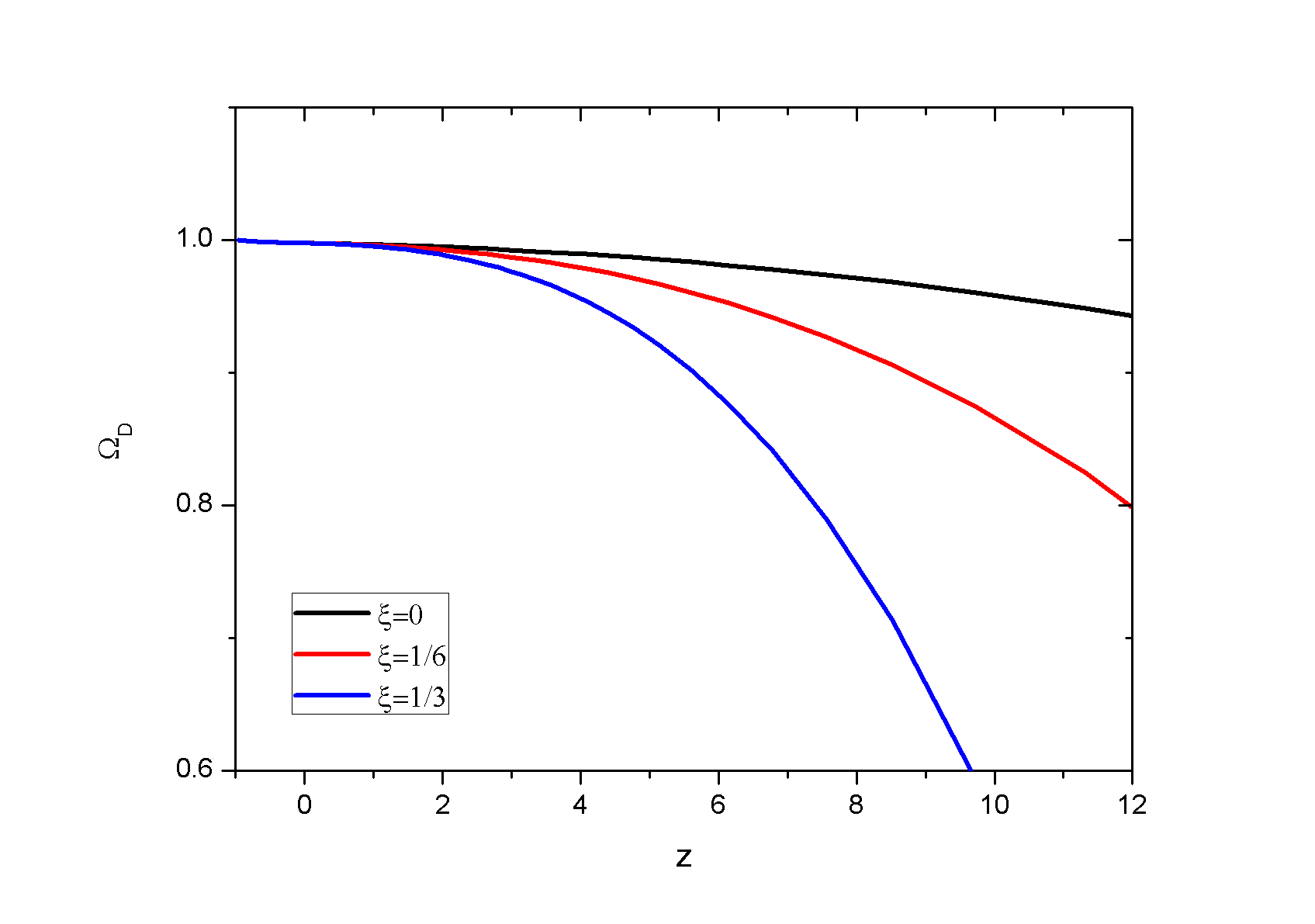}
\caption{DE density parameter as a function of redshift for three representative values of $\xi$ for $k=1.0001633$.}
%\endminipage
\end{figure}

One can note from eqs. \eqref{eq:23}-\eqref{eq:25} that, the anisotropic DE pressure with the skewness parameters depend on the evolutionary behaviour of the functional $\frac{F(R)}{\rho_D}$ and $\frac{\lambda}{\rho_D}$. In the absence of string phase, the pressure anisotropies depend only on the behaviour of $\frac{F(R)}{\rho_D}$. In Fig.4, the evolutionary behaviour of the skewness parameters $\delta, \gamma$ and $\eta$ are shown either in presence or in absence of cosmic strings. One can note that, in absence of cosmic strings (dotted lines in the figure), the skewness parameters show almost a non evolving behaviour in most of the past epochs and evolve at late phase of cosmic time.  $\delta$ and $\eta$ assume negative values whereas $\gamma$ assumes positive values during the evolutionary process. The anisotropy in the DE pressure along x-direction is almost unaffected by cosmic expansion. But the DE pressure along y-direction shows an increasing trend and the DE pressure along z-direction shows a decreasing trend at late times. In presence of cosmic strings, the behaviours of the pressure anisotropies remain almost unaltered at late phase of evolution whereas at an early epoch, the cosmic strings show their presence in a dominant manner. As a result, $\delta$ increases as we move back to past and it changes sign from negative to positive values at some redshift. $\gamma$ also show a similar increasing behaviour with an increase in redshift. However, $\delta$ is more affected by the presence of cosmic strings compared to $\gamma$ and $\eta$. The DE pressure along z-direction is least affected. The reason behind the sensitivity of $\delta$ to the presence of cosmic strings may be due to the fact that, we have considered the strings to align along x-direction. We have also investigated the effect of anisotropy parameter $k$ on pressure anisotropies. It is found that, a variation of $k$, in the given range, does not have any significant effect on the overall behaviour of the skewness parameters.
\begin{figure}[h]
%\minipage{0.40\textwidth}
\includegraphics[scale=0.45]{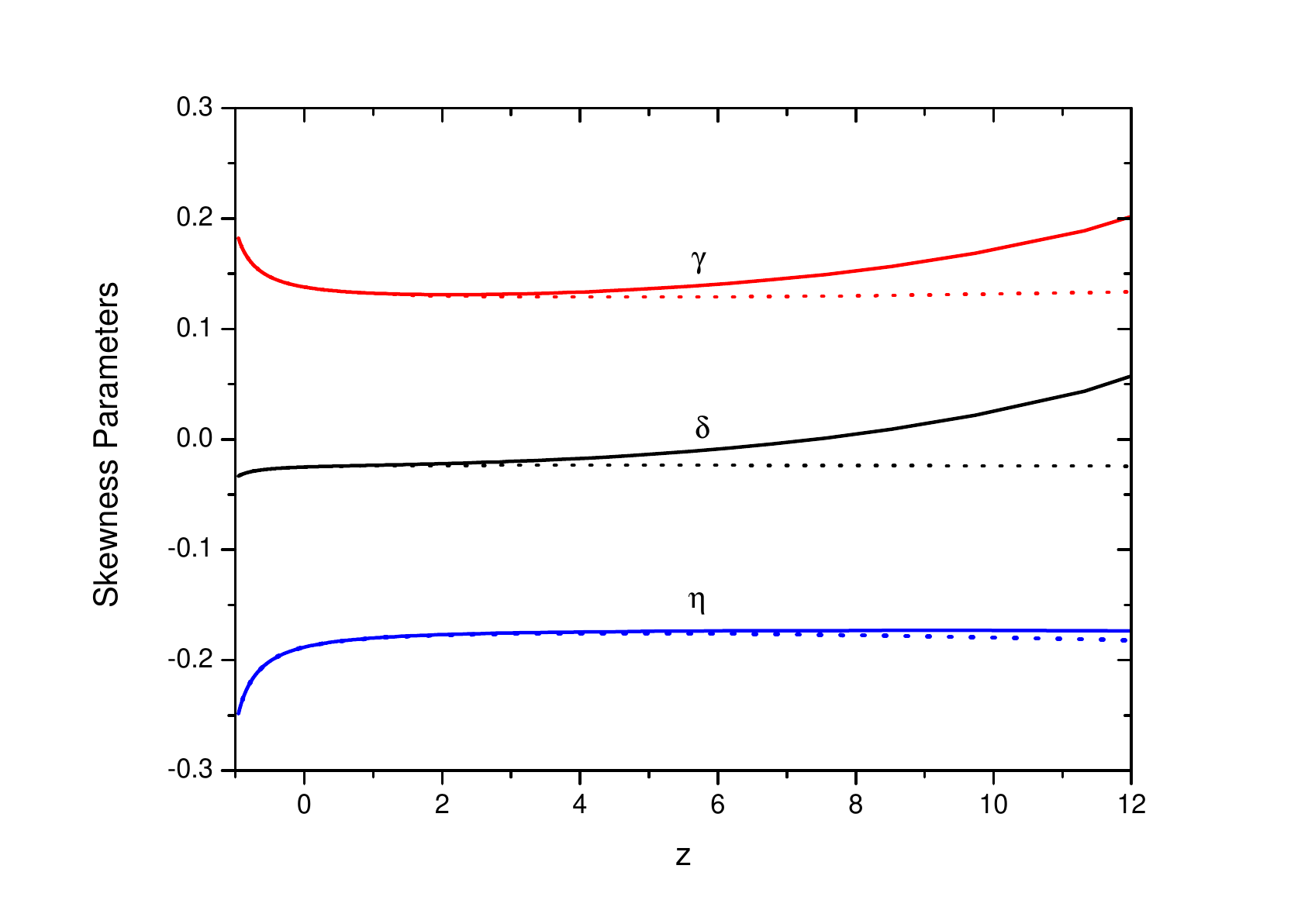}
\caption{Evolution of skewness parameters in presence of cosmic strings. The dotted lines represent the behaviour of skewness parameters in the absence of cosmic strings.}
%\endminipage
\end{figure}

\begin{figure}[h!]
%\minipage{0.40\textwidth}
\includegraphics[scale=0.5]{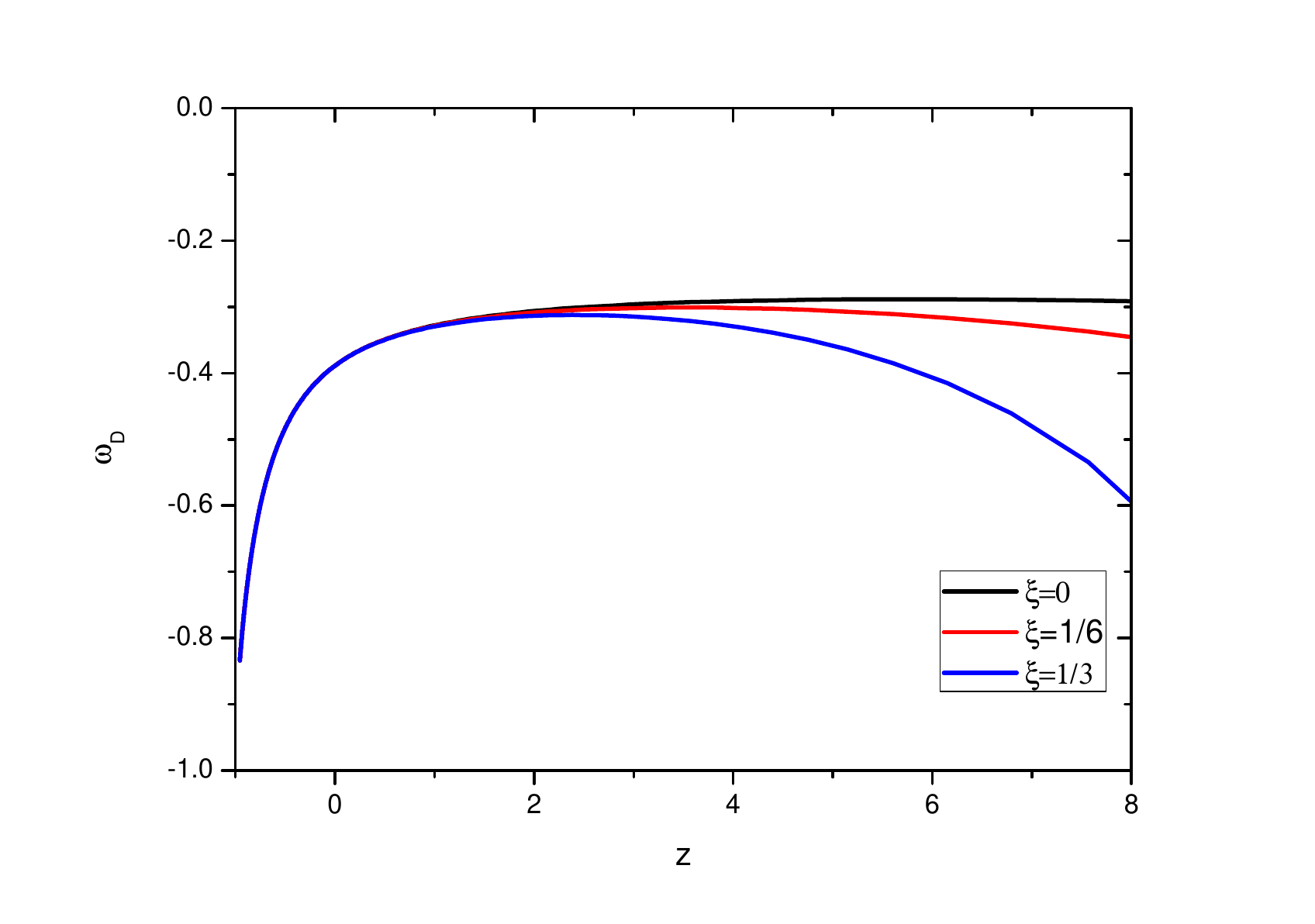}
\caption{Evolution of DE EoS parameter for  three representative strengths of cosmic strings.}
%\endminipage
\end{figure}

\subsection{DE equation of state}
The dynamics of the universe is studied through the evolution of the DE EoS parameter $\omega_D$ in Fig.5.  $\omega_D$ is not sensitive to the choice of the parameter $k$ at least at cosmic times spanning from some recent past to late phase. However, at early times the choice of $k$ may weakly affect the dynamics. In view of this, in the figure we have considered a representative value $k=1.6$ and examined the dynamics in presence of cosmic strings. It is obvious that, at a late phase of cosmic expansion, the dark energy strongly dominates even in the presence of cosmic strings and therefore, the DE EoS is not affected by string anisotropy. But at an early phase, the cosmic string has a substantial contribution to the density parameter and therefore, the dynamics of DE EoS is greatly affected. In the absence of any cosmic strings, the model behaves like quintessence field and the DE EoS smoothly decreases in the quintessence region. With an increase in string tension density, the model gathers some amount of energy in the early phase to behave differently. In such cases, the DE EoS first increases with cosmic time and then decreases.

\section{Conclusion}

In the present work, we have investigated the role of anisotropic components on the dynamical aspects of dark energy. We considered the dark energy to have anisotropic pressures along different spatial directions. Some anisotropic effects have also been incorporated through the inclusion of cosmic strings aligned along x-direction. A hybrid scale factor having two adjustable parameters have been considered to mimic a cosmic transition from an early decelerating phase to a late time acceleration. The parameters of the hybrid scale factor have been fixed from some physical basis. The anisotropic parameter has been tightly bound from recent observations by different authors. In the present work, we have formulated an anisotropic model, where any amount of cosmic anisotropy can be considered and accordingly, the dynamical features of the model can be assessed. However, we have considered that $m=1.0001633$ corresponding to $\left(\frac{\sigma}{H}\right)_0 = 8.164 \times 10^{-5}$ to get some appreciable effects on the dynamical features of the universe. An increase in the anisotropic parameter decreases the DE density parameter and therefore shares some of the burden to provide an acceleration. Presence of string phase substantially modifies the DE density at early epoch. The anisotropy in DE pressure continues along with the cosmic expansion and grows a bit rapidly at a later period. In general,  the pressure anisotropies along different directions increase in the early phase dominated by cosmic strings. Along the x-direction, the anisotropic effect in DE pressure is more due to the string contribution. Along the z-direction, the presence of string phase has little affect on the pressure anisotropy. The DE EoS is also affected due to the anisotropic affect of cosmic strings at early phase. However, the dominance of DE at late times prevents the DE EoS to be affected substantially by cosmic strings.    
 
\section{Acknowledgement}
BM and SKT acknowledge the hospitality of IUCAA, Pune during an academic visit where a part of this work is done.

\end{document}